\documentclass[reprint,aps,superscriptaddress,amsmath,amssymb,longbibliography,nofootinbib]{revtex4-1}
                                                                                                                                                                                                                                                                                                                                                                                                                                                                                                                                                                                                                                                                                                                          
\usepackage{graphicx}
\usepackage{dcolumn}
\usepackage{bm}
\usepackage{color}
\usepackage{xspace}
\usepackage{ulem}
\usepackage{xfrac}
\usepackage{soul}
\usepackage[colorlinks=true,urlcolor=blue,citecolor=blue,linkcolor=blue,bookmarks=false,
pdfstartview={FitH}]{hyperref}

\def \cm-1{cm$^{-1}$}
                                                                                                                                                                                                                                                                                                                                                                                                                                    
\begin{document} 
                                                                                                    
\title{Phonon anomalies within the polar charge density wave phase
of the structurally chiral superconductor Mo$_3$Al$_2$C}

\author{Shangfei Wu}
\email{wusf@baqis.ac.cn}
\affiliation{Department of Physics and Astronomy, Rutgers University, Piscataway, New Jersey 08854, USA}
\affiliation{Beijing Academy of Quantum Information Sciences, Beijing 100193, China}
\author{Xianghan~Xu}
\affiliation{Department of Physics and Astronomy, Rutgers University, Piscataway, New Jersey 08854, USA}
\affiliation{Keck Center for Quantum Magnetism, Rutgers University, Piscataway, New Jersey 08854, USA}
\affiliation{School of Physics and Astronomy, University of Minnesota, Minneapolis, MN, USA}
\author{Fei-Ting~Huang}
\affiliation{Department of Physics and Astronomy, Rutgers University, Piscataway, New Jersey 08854, USA}
\affiliation{Keck Center for Quantum Magnetism, Rutgers University, Piscataway, New Jersey 08854, USA}
\author{Turan~Birol} 
\affiliation{Department of Chemical Engineering and Materials Science, University of Minnesota, MN 55455, USA}
\author{Sang-Wook~Cheong} 
\affiliation{Department of Physics and Astronomy, Rutgers University, Piscataway, New Jersey 08854, USA}
\affiliation{Keck Center for Quantum Magnetism, Rutgers University, Piscataway, New Jersey 08854, USA}
\author{Girsh~Blumberg} 
\email{girsh@physics.rutgers.edu}
\affiliation{Department of Physics and Astronomy, Rutgers University,
Piscataway, New Jersey 08854, USA}
\affiliation{National Institute of Chemical Physics and Biophysics,
12618 Tallinn, Estonia}

\date{\today}             
                                                                                                                                                                                                                                                                                                                                                                                                                                                                                                                                                                                                                                                                  
\begin{abstract}                                             

We employ polarization-resolved Raman spectroscopy to study the lattice dynamics of the polar charge density wave phase of the superconductor Mo$_3$Al$_2$C with structural chirality. We show the phononic signatures of the charge density wave transition at $T^*$ = 155\,K in Mo$_3$Al$_2$C. 
The detailed temperature dependence of these phonon modes'  frequency, half width at half maximum, and integrated area below $T^*$ reveal anomalies at an intermediate temperature $T' \sim$ 100\,K, especially for the low-energy modes at 130 and 180\,\cm-1. {We discuss the origin of these phonon anomalies within the polar charge density wave phase of Mo$_3$Al$_2$C.}

\end{abstract}
                                                    
                                                                                             
\maketitle
                                                                                                                                                                           
\section{Introduction}             
In 1965, Anderson and Blount first proposed the concept of ferroelectric metals or polar metals ~\cite{Anderson_1965_PhysRevLett}. Although they were proposed long ago and many compounds were predicted~\cite{Benedek_2016_JMCC,Young_2023PRM}, only a few examples have been experimentally verified, as ferroelectricity and metallicity have traditionally been considered to be incompatible~\cite{Zhou_2020_review,Ghosez_2022_review,Bhowal_arxiv2022_review}. 
Hence, considerable attention is being directed towards exploring various unconventional mechanisms to realize  polar metal phases~\cite{Benedek_2016_JMCC,Zhou_2020_review,Ghosez_2022_review,Bhowal_arxiv2022_review}.
The decoupled electron mechanism~\cite{Anderson_1965_PhysRevLett,Puggioni_2014_NC,Kim_2016_nature}, which involves a significant separation between the density of states of polar modes and the Fermi level, results in the polar distortion being decoupled from the itinerant electrons surrounding the Fermi sea~\cite{He_2016_PhysRevB,Shi_2013_NatMat,Lei_2018_NanoLett,Benedek_2011_PhysRevLett}.  Another unconventional mechanism for the coexistence of polarization and conductivity was demonstrated experimentally by the switchable ferroelectric metal WTe$_2$ and 1$T'$-MoTe$_2$~\cite{Fei_2018_nature,MoTe} via the interlayer sliding mechanism, which corresponds to the decoupled space mechanism. A third unconventional mechanism is polar bulking in hyperferroelectric metals  resistant to itinerant electrons, as predicted in hexagonal $ABC$ compounds (LiGaGe type)~\cite{Garrity_2014_PhysRevLett}.
                
In principle, breaking symmetry through charge disproportionation and ordering can lead to the emergence of ferroelectricity. In other words, when a noncentrosymmetric charge modulation occurs, it can result in a potential net electric polarization~\cite{Cheong_2007_review,van_den_Brink_2008_review,Fiebig_2016_review}. This concept was proposed and studied in several \textit{insulating} systems, such as magnetite Fe$_3$O$_4$~\cite{Alexe_2009_Advanced}, bond-centered and site-centered Pr$_x$Ca$_{1-x}$MnO$_3$~\cite{Efremov2004}, mixed-valence compounds~\cite{Alonso1999PhysRevLett,Ikeda2005,Groot_2012_PhysRevLett}, and superlattice systems~\cite{Park_2017_PhysRevLett,Krick_2016_Advanced,Qi_2022PRB,Xu_2023_nanolett}.  
However, the phenomenon of charge ordering or charge density wave (CDW) driven polarization in \textit{metallic} systems has been little studied. 

\begin{figure}[!t] 
\begin{center}
\includegraphics[width=\columnwidth]{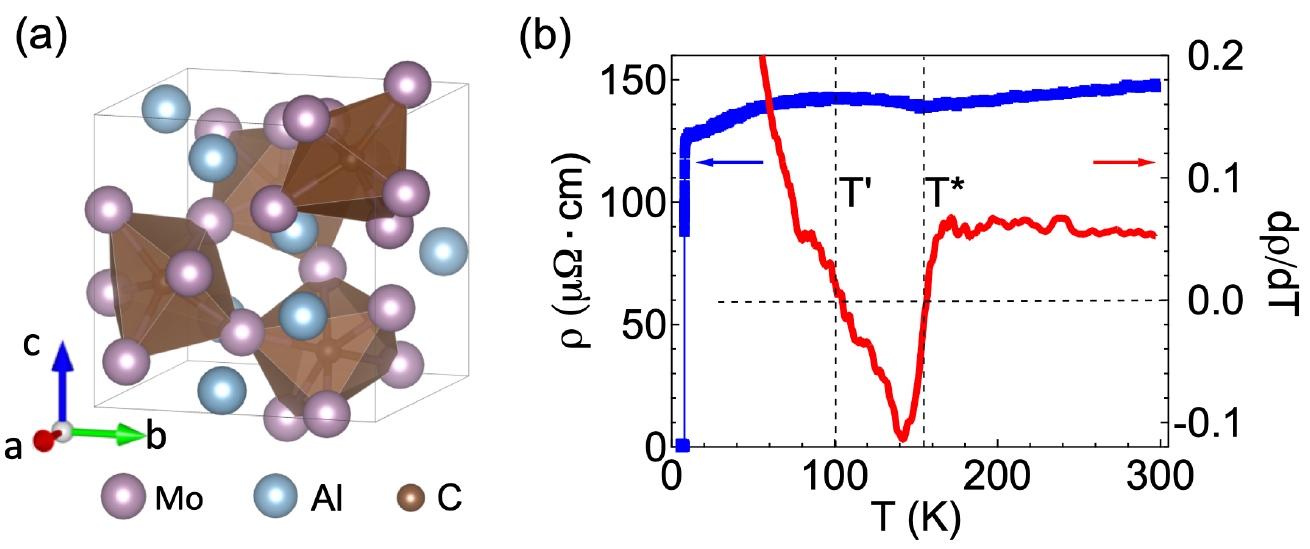}
\end{center}
\caption{\label{Fig1_structure} (a) Crystal structure of Mo$_3$Al$_2$C (space group: $P4_132$ or $P4_332$). (b) Resistivity $\rho$ of Mo$_3$Al$_2$C on the (0~0~1) surface recorded in a cooling-down process for the left axis. The right axis shows the
first derivative of the resistivity as a function of temperature ($\text{d} \rho/\text{d}T$). The two vertical dash lines denote the two anomaly temperatures, $T^*$ = 155\,K and $T' \sim$ 100\,K, corresponding to $\text{d} \rho/\text{d}T=0$.}
\end{figure}

Recently, Mo$_3$Al$_2$C was reported to be a CDW-driven polar metal coexisting with superconductivity and structural chirality~\cite{Wu_2024_NC}. Mo$_3$Al$_2$C has a cubic structure with space group $P4_132$ or $P4_332$ (point group
$O$) at room temperature~[Fig.~\ref{Fig1_structure}(a)]. The crystal structure lacks inversion and mirror symmetries. It is noncentrosymmetric, chiral, and nonpolar at room temperature. 
Upon cooling down the sample, the measured resistivity shows a superconducting transition at $T_c$ = 8\,K and a dip at around $T^*$ = 155\,K~[Fig.~\ref{Fig1_structure}(b)]~\cite{Zhigadlo_2018PRB}. The anomalies at $T^*$ were also reported in previous magnetic susceptibility, specific heat, and nuclear-magnetic-resonance measurements~\cite{Zhigadlo_2018PRB,Koyamado_2013_JPSJ,Koyama2011PhysRevB,Kuo_2012PhysRevB}. Our recent  transmission electron microscopy and Raman scattering results established a CDW order below $T^*$ in Mo$_3$Al$_2$C~\cite{Wu_2024_NC}.  We also showed that the transition at $T^*$ is a cubic-nonpolar to rhombohedral-polar transition driven by condensation of $M_5$ soft modes at three symmetry-equivalent wavevectors located at the Brillouin zone boundaries, creating a polarization along the threefold axis in the body-diagonal direction below $T^*$~\cite{Wu_2024_NC}.
                                                               

In this work, we present a detailed analysis of the lattice dynamics in the polar CDW phase of the superconductor Mo$_3$Al$_2$C using polarization-resolved Raman spectroscopy. We identify the phononic signatures associated with the CDW transition at $T^*$ = 155\,K in Mo$_3$Al$_2$C. An examination of the temperature dependence of the phonon modes’ frequency, half width at half maximum, and integrated area below $T^*$, particularly for the modes at 130 and 180\,\cm-1, reveals anomalies at an intermediate temperature, $T' \sim 100$\,K. We propose that the lattice anomalies at $T'$ within the CDW phase are {possibly} related to a change in the Mo displacements while the crystal symmetry remains unchanged.
    
 \section{Experiment and Methods}\label{ExperimentAndMethods}
                       
\textit{Single-crystal preparation and characterization.}\label{Crystal_preparation}                                                                                                     
Single crystals of Mo$_3$Al$_2$C were grown using a slow cooling method described in Ref.~\cite{Wu_2024_NC}.                                 
Electric transport measurements were carried out using a standard four-point probe method in the (0~0~1) plane in a He exchanges gas environment using a physical property measurement system in a cooling-down process.

\textit{Raman scattering measurements.}\label{Raman} The polished Mo$_3$Al$_2$C crystals with a (0~0~1) plane and (1~1~1) plane used for the Raman scattering study were positioned in
a continuous helium flow optical cryostat~\cite{SM}. 
The Raman measurements
were mainly performed using the Kr$^+$ laser line at 647.1\,nm (1.92\,eV) in
a quasibackscattering geometry along the crystallographic $c$ axis.
The excitation laser beam was focused onto a $50\times100$ $\mu$m$^2$
spot on the $ab$ plane, with the incident power around 12\,mW. 
The scattered light was collected and analyzed by a triple-stage Raman
spectrometer and recorded using a liquid nitrogen-cooled
charge-coupled device. 
Linear and circular polarizations are used in this study to decompose the Raman data into different irreducible representations.
The instrumental resolution was kept better than 1.5\,\cm-1. 
All linewidth data presented in this paper were corrected for  instrumental resolution.
The temperature shown in this paper was corrected for laser heating~\cite{Wu_2024_NC}.

                    
                                       

\textit{Group theory analysis.}\label{TEM} Group theory predictions were performed using the tool provided in the ISOTROPY software suite and the Bilbao Crystallographic Server~\cite{Bilbao_1, Bilbao_4, Hatch2003}.
The information for the irreducible representations of point groups and space groups follows the notation of Cracknell $et~al$.~\cite{cracknell1979general}, which is the same for the Bilbao Crystallographic Server~\cite{Bilbao_4,Bilbao_2}.

 \begin{table}[t]
\caption{\label{SymmetryAnalysis}The relationship between the scattering geometries from a (0~0~1) surface and the symmetry channels for the high-temperature phase. $A_1$, $E$, $T_1$, and $T_2$ are the irreducible representations of the $O$ point group. $X$, $Y$, $X'$, and $Y'$ represent the [1~0~0], [0~1~0], [1~1~0], [1~$-1$~0] directions. $R$ and $L$ represent the right and left circular polarizations defined on the (0~0~1) surface, respectively.}
\begin{ruledtabular}
\begin{tabular}{cc}
Scattering geometry&Symmetry channel (O)\\
\hline
$XX$&$A_1+4E$\\
$XY$&$T_1+T_2$\\
$X'X'$&$A_1+E+T_2$\\
$X'Y'$&$3E+T_1$\\
$RR$&$A_1+E+T_1$\\
$RL$&$3E+T_2$\\
\end{tabular}
\end{ruledtabular}
\end{table}                    

\begin{table}[b]
\caption{\label{decomposition}The algebra used in this study to decompose the Raman data from the (0~0~1) surface into four irreducible representations of point group $O$.}
\begin{ruledtabular}
\begin{tabular}{cc}
Symmetry channel&Expression\\
\hline
$A_1$&$(1/3)(XX+X'X'+RR-X'Y'-RL)$\\
$E$&$(1/6)(X'Y'+RL-XY)$\\
$T_1$&$(1/2)(XY+RR-X'X')$\\
$T_2$&$(1/2)(XY+RL-X'Y')$\\
\end{tabular}
\end{ruledtabular}
\end{table}

\begin{table}[b]
\caption{\label{SymmetryAnalysis22}
The relationship between the scattering geometries from a (1~1~1) surface and the symmetry channels for the high-temperature phase. $A_1$, $E$, $T_1$, and $T_2$ are the irreducible representations of the $O$ point group. 
The selection rules on the (1~1~1) surface are independent of the in-plane angle rotations. Examples of the $X$ and $Y$ directions are the [1~$-1$~0] and  [1~1~$-2$] directions. $R$ and $L$ represent the right and left circular polarizations defined on the (1~1~1) surface, respectively.
}
\begin{ruledtabular}
\begin{tabular}{cc}
Scattering geometry&Symmetry channel ($O$)\\
\hline
$XX$&$A_1+E+T_2$\\
$XY$&$E+T_1+\frac{2}{3}T_2$\\
$RR$&$A_1+T_1+\frac{1}{3}T_2$\\
$RL$&$2E+\frac{4}{3}T_2$\\
\end{tabular}
\end{ruledtabular}
\end{table}

\section{Results and Discussion}

\begin{figure}[!t] 
\begin{center}
\includegraphics[width=\columnwidth]{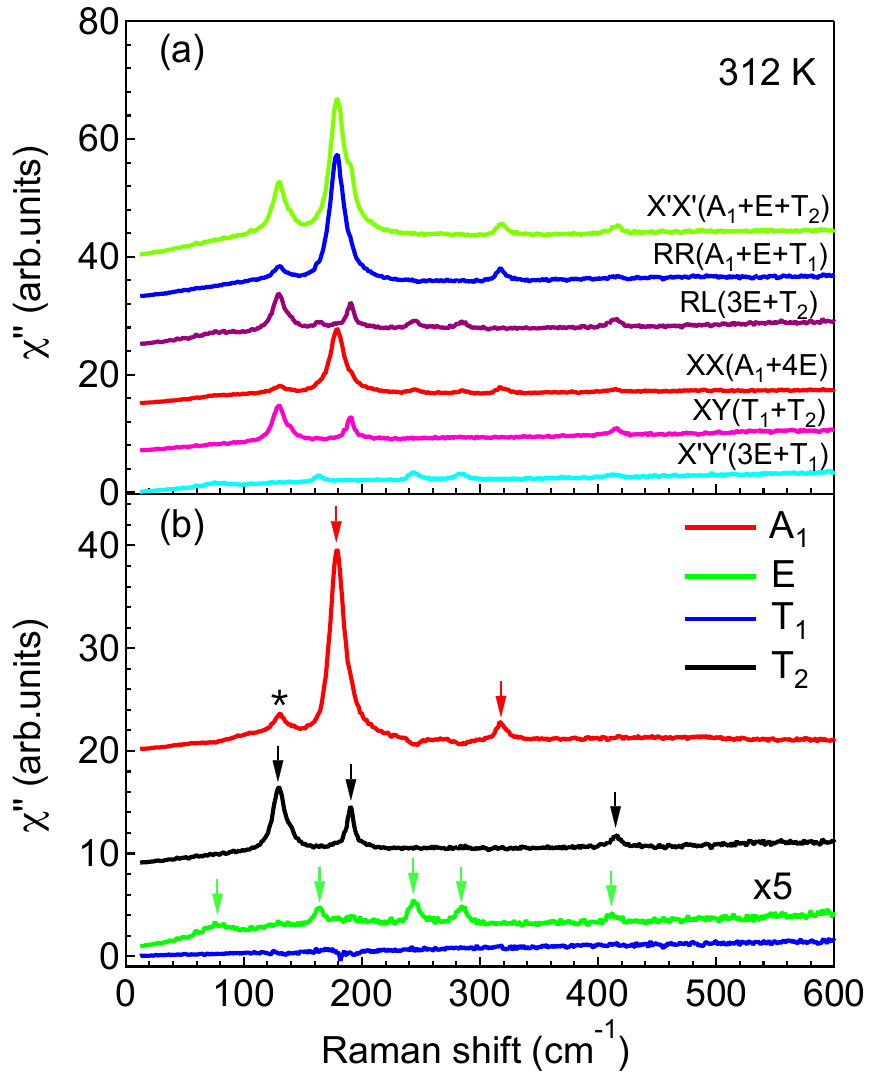}
\end{center}
\caption{\label{Fig2_phonon_300K} (a) Raman spectra of Mo$_3$Al$_2$C from a polished (0~0~1) surface for the $XX$, $XY$, $X'X'$, $X'Y'$, $RR$, and $RL$ scattering geometries at 312\,K. (b) Symmetry decompositions into separate irreducible representations according to the point group $O$ using the algebra shown in Table~\ref{decomposition}. The star in (b) represents the leakage phonon intensity from the $T_2$ phonon at 129\,\cm-1.}
\end{figure}

\begin{figure*}[!t] 
\begin{center}
\includegraphics[width=1.8\columnwidth]{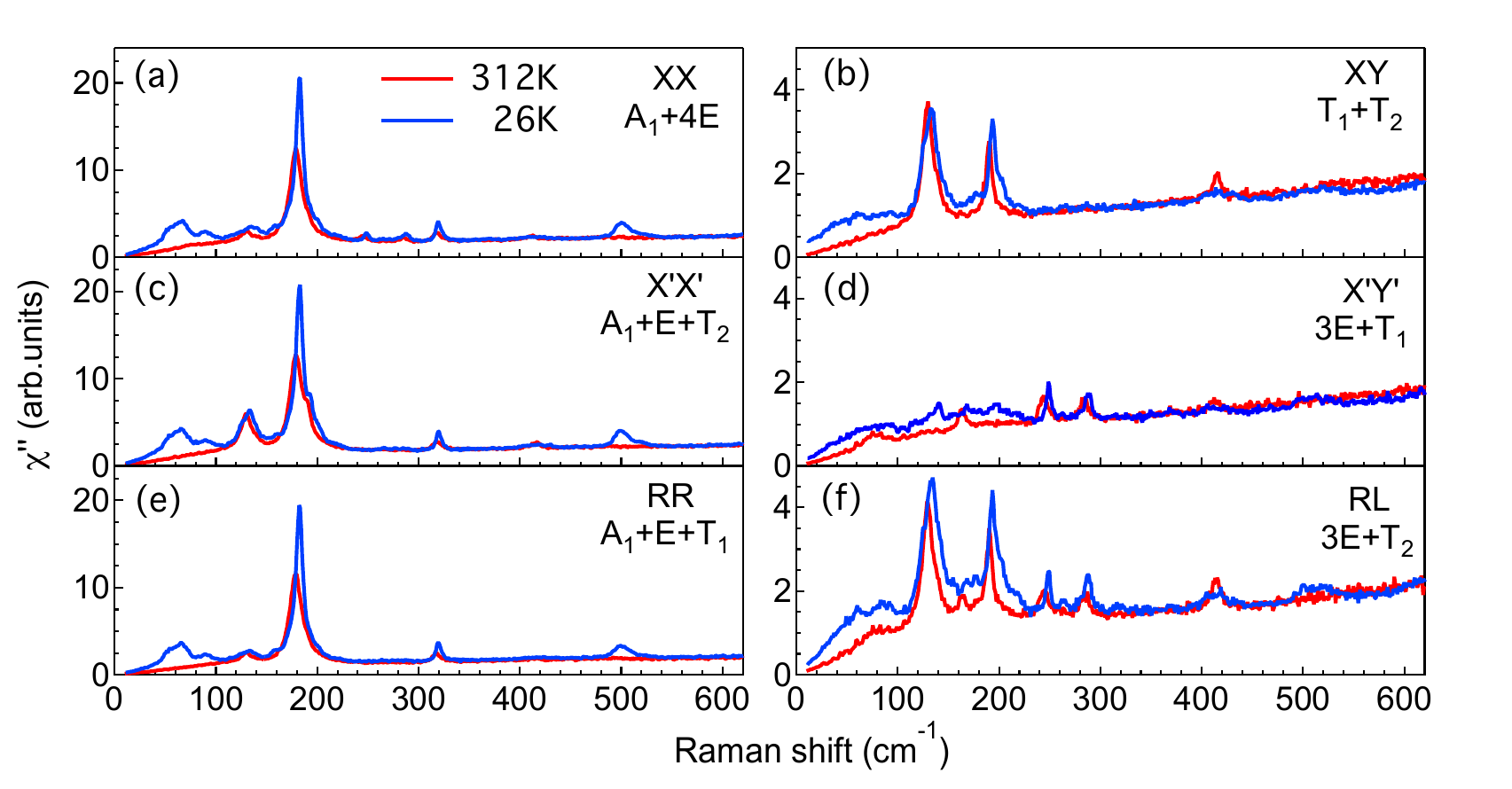}
\end{center}
\caption{\label{compare_300K_10K_chi} Raman spectra of Mo$_3$Al$_2$C on a polished (0~0~1) surface for the (a) $XX$, (b) $XY$, (c) $X'X'$, (d) $X'Y'$ , (e) $RR$, and (f) $RL$ scattering geometries at 312 and 26\,K.}
\end{figure*}

\begin{figure}[!t] 
\begin{center}
\includegraphics[width=\columnwidth]{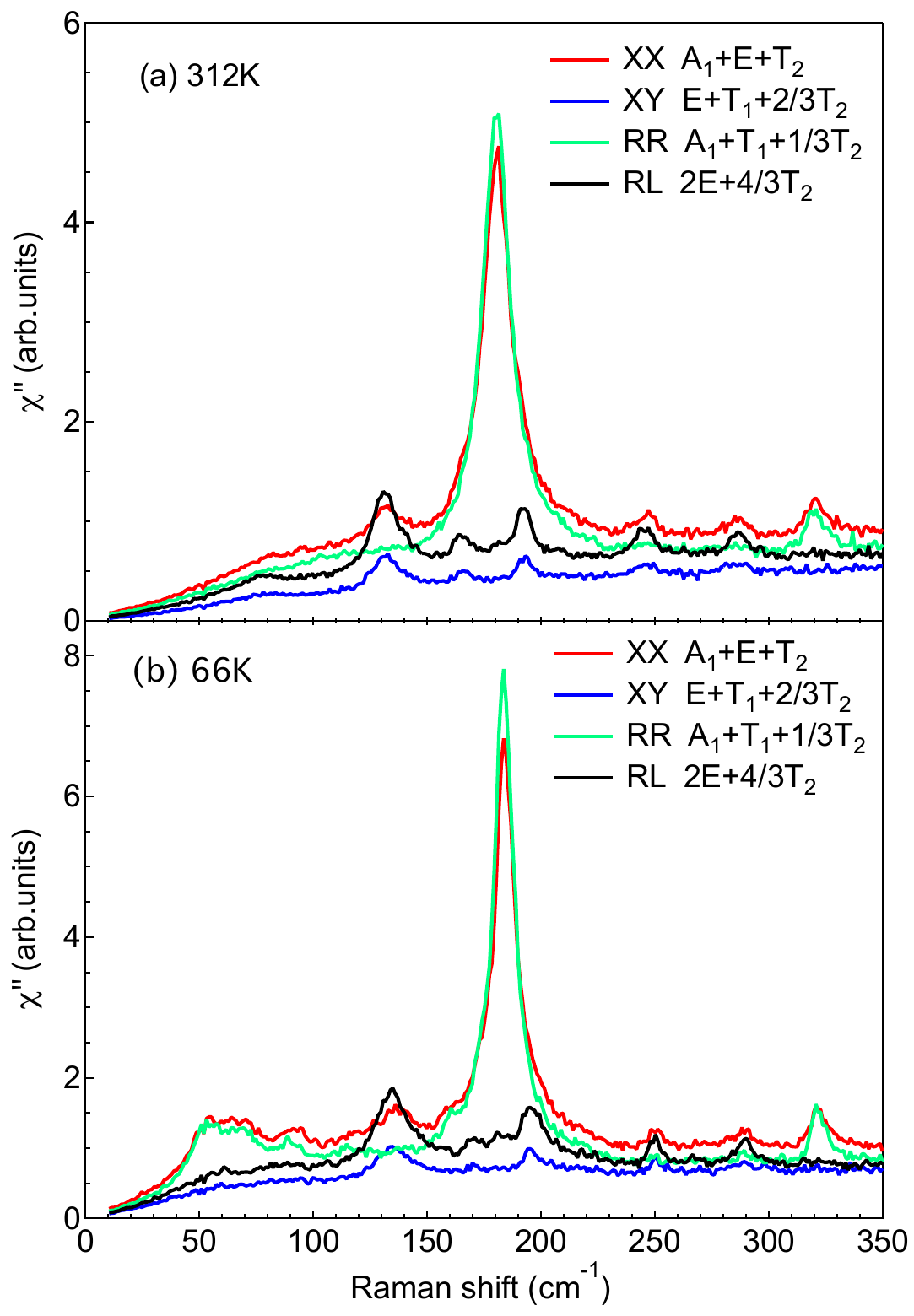}
\end{center}
\caption{\label{Mo3Al2C_111_data} 
{Raman spectra of Mo$_3$Al$_2$C from a polished (1~1~1) surface for the $XX$, $XY$, $RR$, and $RL$ scattering geometries at (a) 312\,K and (b) 66\,K.}
}
\end{figure}

\subsection{Phonon modes}      
The high-temperature phase of Mo$_3$Al$_2$C ($T>T^*$) belongs to a cubic structure with space group $P4_132$ or $P4_332$ [point group
$O$(432)] at room temperature [Fig.~\ref{Fig1_structure}(a)]. One of the phases ($P4_132$ or $P4_332$) is dominant in the sample we studied~\cite{Wu_2024_NC}.
The Mo, Al, and C atoms have Wyckoff positions of $12d$, $8c$, and $4b$, respectively.                                                             
From the group theoretical considerations, $\Gamma$ point phonon modes of  cubic Mo$_3$Al$_2$C can be expressed as $\Gamma_\text{tot}$ = 2$A_{1}$ + 4$A_{2}$ + 6$E$ + 8$T_{2}$ + 10$T_1$. Raman active modes $\Gamma_{\text{Raman}}$= 2$A_{1}$ + 6$E$ + 8$T_{2}$, and IR active modes are $\Gamma_{\text{IR}}$=9$T_1$. Note that $T_1$ signals might become Raman active under resonant conditions~\cite{Placzek1934}. The acoustic mode is $\Gamma_{\text{acoustic}}$ = $T_1$.
Here, we use the notation in the high-temperature phase for the discussion of the phonon modes and their temperature dependence.

For the $O$ point group, the Raman selection rules indicate that the $XX$, $XY$, $X'X'$, $X'Y'$, $RR$, and $RL$ polarization geometries probe the $A_1 + 4E$, $T_1+ T_2$, $A_1 + E + T_2$, $3 E + T_1$, $A_1 + E + T_1$, and $3 E + T_2$ symmetry excitations, respectively. Symmetry analysis for  point group $O$ is summarized in Table~\ref{SymmetryAnalysis}~\cite{SM}. The algebra used to decompose measured Raman spectra into four irreducible representations for point group $O$ are shown in Table~\ref{decomposition}~\cite{SM}.
                                        
In Fig.~\ref{Fig2_phonon_300K}(a), we show the phonon spectra of Mo$_3$Al$_2$C at room temperature in the six measured scattering geometries for a polished (0~0~1) surface. Several phonon modes with $A_1$, $E$, and $T_2$ symmetries are observed in these scattering geometries. As we show the symmetry decompositions according to point group $O$ (Table~\ref{decomposition}) in Fig.~\ref{Fig2_phonon_300K}(b), these phonon modes are separated into the $A_1$, $E$, $T_1$, and $T_2$ symmetry channels.                  
In the $A_1$ symmetry channel, we detect two phonon modes at 180 and 319\,\cm-1, corresponding to Mo and Al fully symmetric lattice vibrations, respectively. In the $T_2$ symmetry channel, we detect three phonons at 129, 190, and 415\,\cm-1. Note that the low-energy $T_2$ at 129\,\cm-1 leaks into the $A_1$ symmetry channel, creating a weak peak in the $A_1$ symmetry channel. This leakage signal might be due to imperfect cutting and polishing of the single crystal. 
In the $E$ symmetry channel, we observe five phonon modes at 75, 164, 245, 285, and 413\,\cm-1. These $E$-symmetry phonon intensities are generally 5 times weaker than the $T_2$-symmetry phonons. The four sharp $E$ modes are first-order phonons because the broad one at 75\,\cm-1 is a second-order two-phonon excitation as the linewidth is 4 times that of the three sharp ones.
In the $T_1$ antisymmetric channel, we barely detect any phonon modes. The peak positions of the $A_1$, $E$, and $T_2$ phonons are summarized in Table~\ref{table_phonon}.

\begin{table}[b]
\caption{\label{table_phonon} Summary of experimental phonon mode  symmetries and frequencies for Mo$_3$Al$_2$C at 312\,K.} 
\begin{ruledtabular}
\begin{tabular}{cc}
Symmetry&Expt.(\cm-1)\\
\hline
$E$&75\\
$T_2$&129 \\
$E$&164\\
$A_1$&180\\
$T_2$&190\\
$E$&245\\
$E$&285\\
$A_1$&319\\
$E$&413\\
$T_2$&415\\
\end{tabular}
\end{ruledtabular}
\end{table}  

\begin{figure*}[!ht] 
\begin{center}
\includegraphics[width=2\columnwidth]{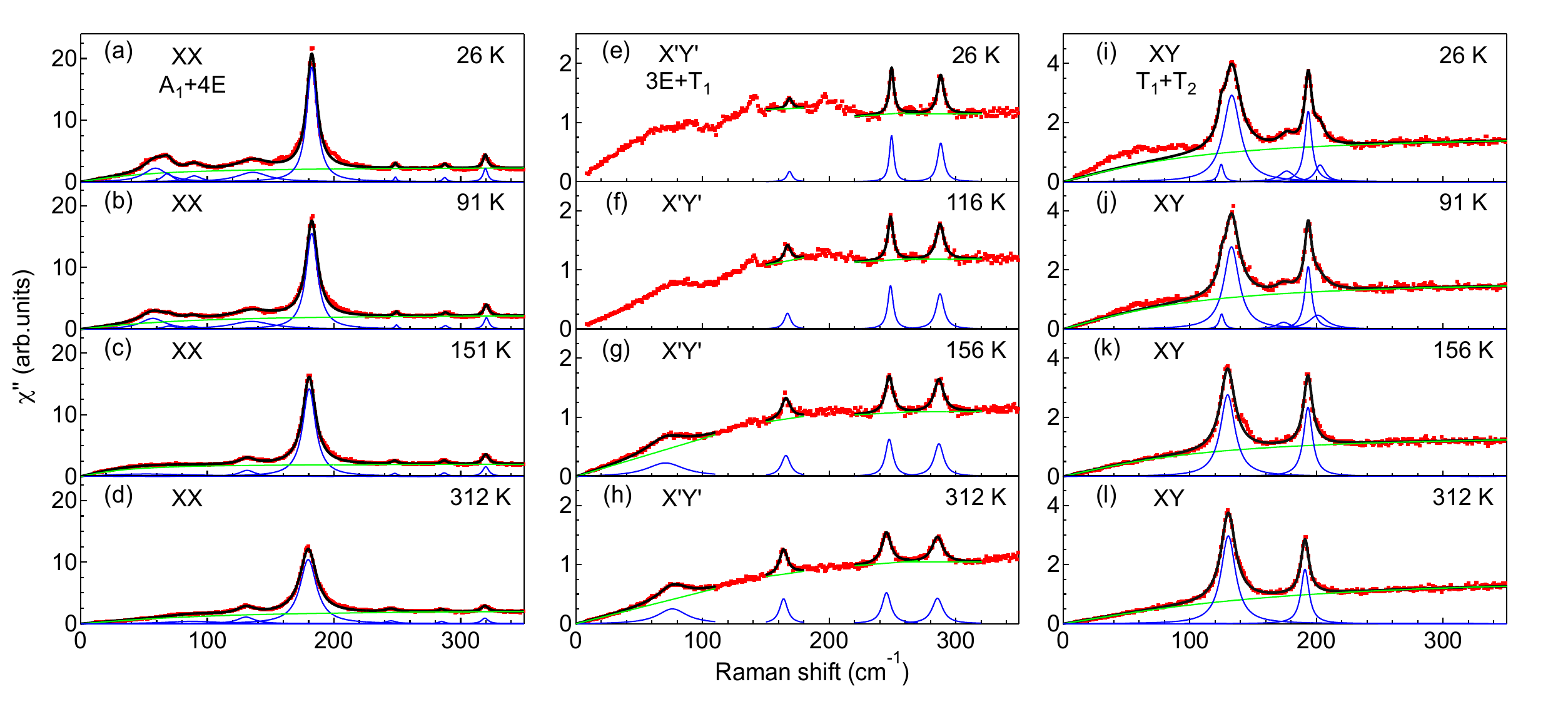}
\end{center}
\caption{\label{Fitting}   
Selected fitting results in the $XX$, $X'Y'$, and $XY$ scattering geometries on the (0~0~1) plane of Mo$_3$Al$_2$C at 312, 151, 91, and 26\,K using Lorentzian function fitting on a smooth background. 
(a)-(d) are for the $XX$ scattering geometry, (e)-(h) are for the $X'Y'$ scattering geometry, and (i)-(l) are for the $XY$ scattering geometry.
The black solid curves denote the total response. The green curves denote the backgrounds. The blue curves denote the individual components.}
\end{figure*}  

Upon cooling below $T^* = $155\,K, Mo$_3$Al$_2$C undergoes a CDW transition and becomes  a rhombohedral $R3$ structure with point group $C_3$. The primitive cell at low temperature becomes a $2\times 2 \times1$ superstructure and thus 4 times larger in the hexagonal setting~\cite{Wu_2024_NC}. The correlation table for point group $O$ indicates that the $A_1$ and $A_2$ irreducible representations merge into the $A$ irreducible representation in the $R3$ phase, the $E$ irreducible representation does not change, and the $T_2$ irreducible representation splits into $A+E$ in the $R3$ phase~\cite{Wu_2024_NC}.
From the group theoretical considerations, the $\Gamma$ point phonon modes of the rhombohedral Mo$_3$Al$_2$C can be expressed as $\Gamma_\text{tot}$ = 96$A$ $\oplus$ 96$E$. Raman active modes and IR active modes are $\Gamma_{\text{Raman}}$ = $\Gamma_{\text{IR}}$ = 95$A \oplus 95E$, and the acoustic mode $\Gamma_{\text{acoustic}}$ =$A \oplus E$. 
Note that the fully symmetric carbon lattice vibration mode becomes Raman active in the $C_3$ phase below $T^*$. 
                              
As shown in Figs.~\ref{compare_300K_10K_chi}(a),~\ref{compare_300K_10K_chi}(c), and~\ref{compare_300K_10K_chi}(e), several low-energy $A$ symmetry modes appear below 100\,\cm-1 in the parallel scattering geometries below $T^*$. They correspond to the Mo lattice vibration modes according to the DFT phonon calculation~\cite{Reith_2012_PhysRevB}.  Specifically, the broad peak at around 50\,\cm-1 appearing at low temperatures is the amplitude mode of the CDW order parameter~\cite{Wu_2024_NC}.
Furthermore, a noticeable $A$ symmetry peak at around 500\,\cm-1 appears in the $XX$, $X'X'$, and $RR$ scattering geometries below $T^*$. A weaker $E$-symmetry peak at around 520\,\cm-1 appears in the $X'Y'$ and $RL$ scattering geometries. These modes correspond to the carbon lattice vibrations modes~\cite{Reith_2012_PhysRevB}, which appear only in the symmetry-broken state according to group-theoretical analysis.
Last, below $T^*$, the four first-order $E$ symmetry phonon modes do not split because they can be fitted by a single Lorentzian function, as shown in Figs.~\ref{Fitting}(e)-~\ref{Fitting}(h).
In the $XY$ scattering geometry, we detect two $T_2$ symmetry phonons at 129 and 190\,\cm-1 at 312\,K. They split into two modes in the 26\,K data as shown in Fig.~\ref{Fitting}(i)-\ref{Fitting}(l).

In Fig. 4, we present the Raman response of Mo$_3$Al$_2$C at 312 and 66\,K in the four measured scattering geometries from a polished (1~1~1) surface. The phonon spectra at 312 and 66\,K are consistent with the Raman data measured from the (0~0~1) surface (Figs.~\ref{Fig2_phonon_300K} and~\ref{compare_300K_10K_chi}). Regarding the selection rule for the (1~1~1) surface~(Table~\ref{SymmetryAnalysis22})~\cite{SM}, the $A_1$ and $E$ irreducible representations can be separated in the $RR$ and $RL$ scattering geometries, respectively. This is impossible for the Raman response from the (0~0~1) surface because the phonons with $A_1$ and $E$ symmetries always appear together in the parallel scattering geometries. This point is rather helpful in the discussion of the second-order Raman response in the next section.

\subsection{Temperature dependence}\label{phonon_modes}  

\begin{figure*}[t] 
\begin{center}
\includegraphics[width=2\columnwidth]{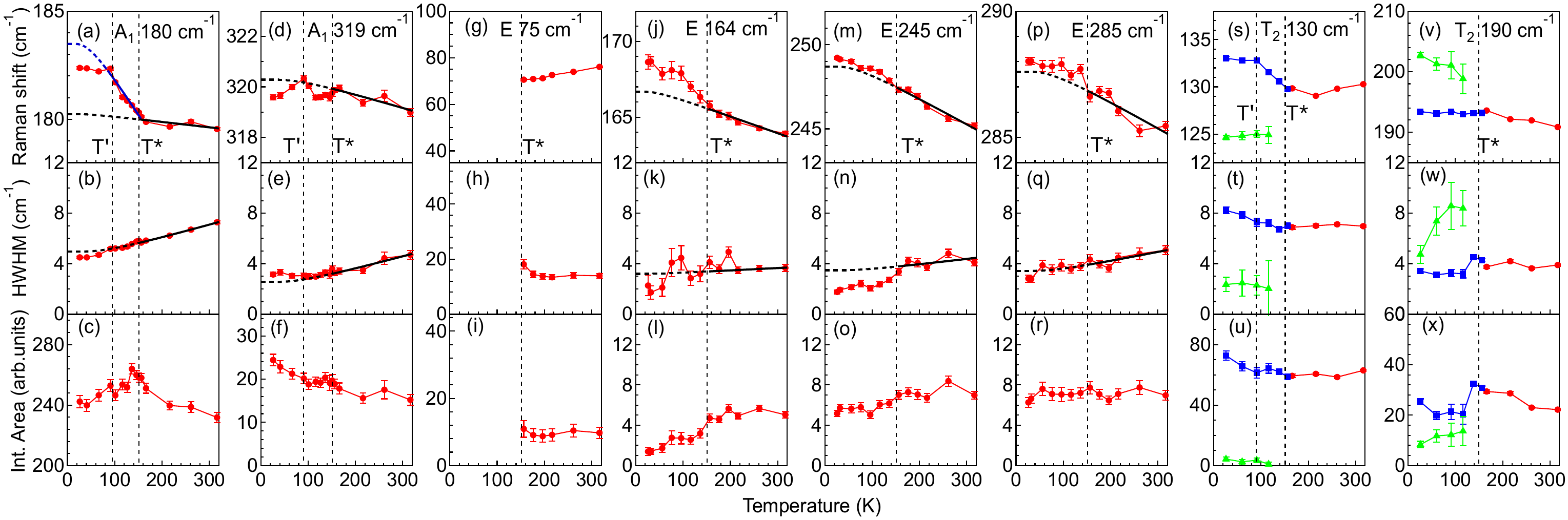}
\end{center}
\caption{\label{Fitting_parameters} T dependence of the fitting parameters [peak frequency, half width at half maximum (HWHM), integrated area] for the $A_1$, $E$ and $T_2$ phonon modes. (a)-(c) are for $A_1$ mode at 180\,\cm-1, {(d)-(f) are for $A_1$ mode at 319}\,\cm-1, (g)-(i) are for $E$ mode at 75\,\cm-1, (j)-(l) are for $E$ mode at 164\,\cm-1, (m)-(o) are for $E$ mode at 245\,\cm-1, (p)-(r) are for $E$ mode at 285\,\cm-1, (s)-(u) are for $T_2$ mode at 130\,\cm-1, and (v)-(x) are for $T_2$ mode at 190\,\cm-1. The dashed vertical lines represent $T^*$ at 155\,K and $T'$ at around 100\,K.
}
\end{figure*}

After establishing the phonon modes in Mo$_3$Al$_2$C, we switch to studying their temperature dependence.                                                                                                                             

In Fig.~\ref{Fitting}, we show the fitting results for data in the $XX$, $X'Y'$, and $XY$ scattering geometries from the (0~0~1)  {surface} of Mo$_3$Al$_2$C at selected temperatures (312, 151, 91, and 26\,K) using multi-Lorentzian peaks fitted on a smooth background~\cite{SM}.
In Fig.~\ref{Fitting_parameters}, we present the temperature dependence of the peak frequencies, half width at half maximum (HWHM), and intergraded area for these $A_1$, $E$, and $T_2$ phonon modes.

As shown in Fig.~\ref{Fitting_parameters}(a), the $A_1$ phonon mode at 180\,\cm-1 hardens upon cooling. It 
shows substantial additional hardening upon cooling below $T^*$, suggesting this mode couples to the CDW order parameter. 
Below $T'  \sim 100$\,K, the $A_1$ phonon mode at 180\,\cm-1 shows a saturation-like behavior. The anomaly at $T'$ is also found in the $T$-dependence of the resistivity curve where  the slope changes sign at $T'$, namely, $\text{d} \rho/\text{d}T$ is zero at around $T'$~[Fig.~\ref{Fig1_structure}(b)].
The temperature dependence of the mode frequency follows well the anharmonic phonon decay model~(Appendix~\ref{Anharmonic_decay_model}) 
in both temperature ranges $T>T^*$ and $T' <T<T^*$. 
On the contrary, the $A_1$ mode at 319\,\cm-1 decreases a little bit upon cooling below $T^*$, then increases, finally decreases again, and forms a local maximum at $T' $.
For the HWHMs of these two modes, both of them decease upon cooling, showing anomalies at $T^*$.
For the integrated area, the $A_1$ phonon mode at 180\,\cm-1 shows an increase upon cooling and decreases below $T^*$. In contrast, the $A_1$ phonon mode at 319\,\cm-1 increases continuously upon cooling, indicating the finite electron-phonon coupling in this system~\cite{Mai2019PRB}. {We note that the $E$ mode at 140\,\cm-1 in the $X'Y'$ scattering geometry [Figs.~\ref{compare_300K_10K_chi}(d), ~\ref{Fitting}(e), and \ref{Fitting}(f)], which appears only below $T^*$, shows the Fano line shape reported in previous studies~\cite{Xu_2017_NC,Wu2020PhysRevB}. The asymmetric line shape for this mode also supports the existence of electron-phonon coupling in this system.}
                                                                                                                                                                                                                                                                                                                                                                                                                                                                                                                                                                                                                                                                                             
For the $E$ modes at 164, 245, and 285\,\cm-1, these three phonons harden upon cooling, showing additional hardening below $T^*$. 
The additional hardening is larger for the mode at 164\,\cm-1 than the other two modes~[Figs.~\ref{Fitting_parameters}(j), \ref{Fitting_parameters}(m), and~\ref{Fitting_parameters}(p)].
These three phonon modes narrow upon cooling. The HWHMs for the modes at 164 and 285\,\cm-1 follow well the anharmonic phonon decay model~[Figs.~\ref{Fitting_parameters}(k) and~\ref{Fitting_parameters}(q)], while the mode at 245\,\cm-1 narrows much faster below $T^*$~[Fig.~\ref{Fitting_parameters}(n)].
For the integrated area, the two phonons at 245 and 285\,\cm-1 remain almost constant~[Figs.~\ref{Fitting_parameters}(o) and \ref{Fitting_parameters}(r)], while the integrated area decreases quickly for the mode at 164\,\cm-1 below $T^*$~[Fig.~\ref{Fitting_parameters}(l)].
                                                                        
The $E$ mode at 75\,\cm-1 softens to 70\,\cm-1 from 312\,K to $T^*$~[Fig.~\ref{Fitting_parameters}(g)]. In contrast, the linewidth and the integrated area for this mode increase slightly upon cooling close to $T^*$~[Figs.~\ref{Fitting_parameters}(h) and \ref{Fitting_parameters}(i)]. Note that the linewidth of the 75\,\cm-1 mode is 4 times those of the other $E$ modes. The softening of the 75\,\cm-1 mode above $T^*$, {the spectrum line shape change below $T^*$}, and its large line-width broadening suggest that it is not a regular first-order phonon mode, but rather a second-order phonon that reflects the softening phonon branch  around $M$ points~\cite{Wu_2024_NC}, similar to what was found in the study of $2H$-NbSe$_2$~\cite{mialitsin2010raman}. {The softening of the phonon branch at the $M$ point could be probed by future momentum-resolved spectroscopies, such as electron energy loss spectroscopy or inelastic neutron/x-ray scattering spectroscopy.}
This $E$-type second-order phonon is symmetry allowed based on the direct product rules: $M_5 \otimes M_5=\Gamma_1(A_1)  \oplus \Gamma_3(E) \oplus \Gamma_4(T_1) \oplus \Gamma_5(T_2)$ (Appendix~\ref{Direct_product}). The details of the second-order scattering process are also given in Appendix~\ref{Direct_product}.
{Generally, for second-order Raman scattering, the identity representation $\Gamma_1(A_1)$ produces scattering efficiencies considerably stronger than those of the other channels, as is the case for Si~\cite{CardonaBookII}.
For Mo$_3$Al$_2$C, the second-order Raman scattering intensity is more noticeable in the $E$ symmetry channel [$RL$ in Fig.~\ref{Mo3Al2C_111_data}(a)] than that in the $A_1$ symmetry channel [$RR$ in Fig.~\ref{Mo3Al2C_111_data}(a)], rendering Mo$_3$Al$_2$C a special case for second-order Raman scattering.}

For the $T_2$ phonon modes, these two modes at 130 and 190\,\cm-1 are observed above $T^*$, as shown in Figs.~\ref{Fitting}(i)-\ref{Fitting}(l).
Below $T^*$, the $T_2$ mode splits into two modes, as shoulder peaks can be seen on the left side of the mode at 130\,\cm-1 and on the right side of the mode at 190\,\cm-1.  The splitting of the $T_2$ phonon modes is due to the symmetry breaking as a result of the CDW order below $T^*$~\cite{Wu_2024_NC}.
The splittings can be seen from the fitting of these modes shown in Figs.~\ref{Fitting}(i)-\ref{Fitting}(l). Above $T^*$, a single Lorentzian function can describe the $T_2$ phonon well while it requires two Lorentzian functions to model the $T_2$ phonon below $T^*$.
Note that a separate peak at 177\,\cm-1 appears below $T^*$. It is a new mode, and it does not belong to one of the components of the $T_2$ mode at 190\,\cm-1.

The fitting parameters for the $T_2$ phonon modes are presented in Figs.~\ref{Fitting_parameters}(s)-\ref{Fitting_parameters}(x). For the $T_2$ mode at 130\,\cm-1, it softens 1\,\cm-1 upon cooling from 312 to 200\,K~[Fig.~\ref{Fitting_parameters}(s)], while the HWHM and the integrated area barely change~[Figs.~\ref{Fitting_parameters}(t) and \ref{Fitting_parameters}(u)]. Below $T^*$, it splits into two modes: One is at 125\,\cm-1, and the other one is at 133\,\cm-1. 
For the $T_2$ mode at 190\,\cm-1, it hardens slightly upon cooling~[Fig.~\ref{Fitting_parameters}(v)], while the HWHM barely changes~[Fig.~\ref{Fitting_parameters}(w)] and the integrated area increases slightly~[Fig.~\ref{Fitting_parameters}(x)]. Below $T^*$, it splits into two modes: One is at 193\,\cm-1, and the other one is at 202\,\cm-1. 
We note that the high-energy component of the split 130\,\cm-1 mode shows an anomaly at $T'$; namely, the $T$-dependence of the mode's frequency shows a saturation-like behavior below $T'$~[Fig.~\ref{Fitting_parameters}(s)], similar to the $A_1$ mode at 180\,\cm-1~[Fig.~\ref{Fitting_parameters}(a)].

%
                                                                                                                                                                                                                                                                                        
Finally, we discuss the origin of the phonon and resistivity anomalies
at $T'$. Based on the density functional theory phonon calculations, there are three unstable phonon modes in the phonon dispersion for the cubic phase of Mo$_3$Al$_2$C: $\Gamma_4$ (0,~0,~0), $M_5$ (0.5,~0.5,~0), and $X_2$ (0.5,~0.5,~0)~\cite{Wu_2024_NC,Reith_2012_PhysRevB}.  
In our previous study~\cite{Wu_2024_NC}, we showed that Mo$_3$Al$_2$C undergoes a cubic-nonpolar to rhombohedral-polar transition below $T^*$ into an $R3$ phase, which is driven by condensation of the $M_5$ soft modes at three symmetry equivalent wavevectors located at the Brillouin zone boundary, and creates a polarization along the three-fold axis in the body-diagonal direction below $T^*$. The analysis of the $T$-dependence of the superlattice Bragg peaks and polar domain imaging support that the CDW transition and the polar transition occur at the same temperature $T^*$~\cite{SM}. 
From the $T$ dependence of the Raman response in the $XX$, $XY$, and $X'Y'$ scattering geometries~(Fig.~\ref{Fitting} here and Fig.~3 in Ref.~\cite{Wu_2024_NC}), the number of new phonon modes does not change below $T'$; thus, the phonon anomalies at $T'$ do not seem to be related to zone-boundary lattice instabilities. {Furthermore, the twofold-degenerate $E$ modes at 245 and 285\,\cm-1 do not show signatures of splitting within our energy resolution (1.5\,\cm-1 based on the full-width at half-maximum for the laser line) in the $R3$ phase (Fig.~\ref{compare_300K_10K_chi} and Fig.~\ref{Fitting}). The linewidths for these $E$ modes do not show noticeable broadening upon cooling across $T'$~[Figs.~\ref{Fitting_parameters}(n) and \ref{Fitting_parameters}(q)].}

 {
The deviation from the anharmonic behavior of the $A_1$ mode at 180\,\cm-1 below $T'$ could be due to either lattice distortion, electron-phonon coupling, or spin-phonon coupling. The $A_1$ mode at 180\,\cm-1 shows  a symmetric line shape between 300 and 26\,K, and can be modeled by a Lorentzian line shape~(Fig.~\ref{Fitting}). Thus, the electron-phonon coupling mechanism is not likely. The absence of magnetic order in Mo$_3$Al$_2$C from the magnetic susceptibility measurement~\cite{Koyamado_2013_JPSJ} suggests that spin-phonon coupling could not account for the phonon anomaly at $T'$.}

 {
 One possible explanation for the phonon anomalies at $T'$ could be a fully symmetric modification of the Mo lattice structure while the overall $C_3$ point-group symmetry is preserved, because the Mo lattice related vibration modes~\cite{Reith_2012_PhysRevB}, especially for the $T_2$ mode at 130\,\cm-1 and the $A_1$ mode at 180\,\,\cm-1, show clear saturationlike behaviors below $T'$. The fully symmetric modification of the crystal structure in Mo$_3$Al$_2$C below $T'$ might resemble the case in Na$_{0.5}$CoO$_2$, where the overall space group symmetry for Na$_{0.5}$CoO$_2$ remains $Pnmm$ between 300 and 10\,K~\cite{Huang_2004_JPCM,Williams_2006_PhysRevB} but the Na-O bond lengths show multiple anomalies in the same temperature range according to the neutron powder diffraction structural refinements~\cite{Argyriou_2007_PhysRevB}. We note that the temperature dependence of the amplitude mode at around 50\,\cm-1 in the CDW phase~(Figs.~3(f)-3(h) of Ref.~\cite{Wu_2024_NC}) shows a smooth evolution upon cooling across $T'$. This suggests that the fully symmetric modification of the Mo lattice structure below $T'$ is secondary to the dominant lattice distortion due to the soft mode, which condenses below the CDW transition. This point is also supported by the temperature dependence of the resistivity curve, where a gradual change in the resistivity slope is found at $T'$, in contrast to the clear dip found at $T^*$~[Fig.~\ref{Fig1_structure}(b)]. Thus, the physics for triggering the change in the modulation pattern associated with the Mo lattice across $T'$ is different from what is dictated by the $M_5$ soft mode eigenvectors.
 }
          
 {
While we cannot completely rule out a symmetry-breaking phase transition at $T'$ based on our current energy resolution in the Raman experiment, future high-resolution x-ray structure refinement investigations of the lattice structure below $T^*$ and $T'$ could determine the details of the phase transitions and lattice anomalies in Mo$_3$Al$_2$C.
}

\section{CONCLUSIONS}    
In summary, we studied the lattice dynamics of the polar CDW phase of the superconductor Mo$_3$Al$_2$C using polarization-resolved Raman spectroscopy. 
We investigated the temperature dependence of the phonon modes' frequency, half width at half maximum, and integrated area below $T^* = 155$\,K and found anomalies at an intermediate temperature, $T' \sim$ 100\,K, {for the modes at 130 and 180\,\cm-1}. 
{We inferred that the lattice anomalies at $T'$ are {possibly} related to a  {fully symmetric} modification of Mo displacements within the CDW phase while the lattice symmetry is preserved};
high-resolution x-ray structure refinement investigations of the lattice distortions below $T^*$ and $T'$ are required to further understand the phase transitions  {and lattice anomalies} in the polar CDW compound Mo$_3$Al$_2$C.  
Our results in this paper provide detailed lattice dynamics information for Mo$_3$Al$_2$C, forming a basis for future studies of the superconducting state in Mo$_3$Al$_2$C which processes both structural polarity and chirality, e.g., switchable ferroelectric superconductivity~\cite{Jindal_2023Nature}, nonreciprocal charge transport~\cite{Tokura_2018NC,Nadeem_2023review}, and the unconventional paring mechanism~\cite{Yip_2014Review,Kallin_2016}.

\begin{acknowledgments}
The spectroscopy work at Rutgers (S.F.W and G.B.) was supported by NSF Grants No.~DMR-2105001. 
The work at BAQIS (S.F.W.) was supported by the National Natural Science Foundation of China (Grant No.~12404548).  
The sample growth, characterization, and TEM work (X.H.X., F.T.H, and S.W.C.) was supported by the W. M. Keck foundation grant to the Keck Center for Quantum Magnetism at Rutgers University.
T.B. was supported by the Department of Energy through the University of Minnesota Center for Quantum Materials under DESC0016371.
The work at NICPB was supported by the European Research Council (ERC) under the European Union’s Horizon 2020 research and innovation programme grant agreement No.~885413.  \\                                                                                                                          
\end{acknowledgments}

  
\appendix

\section{Anharmonic phonon decay model}\label{Anharmonic_decay_model}    
In this appendix, we discuss the anharmonic phonon decay model.
We fit the temperature dependence of the phonon frequency and HWHM using the anharmonic phonon decay model~\cite{Klemens_PhysRev148,Cardona_PRB1984}: 
\begin{equation}
\label{eq_omega1}
\omega_1(T)=\omega_{0}-C_1\{1+ 2n(\Omega(T)/2) \},\\
\end{equation}
\begin{equation}
\label{eq_gamma1}
\Gamma_1(T)=\gamma_{0}+\gamma_1\{1+ 2n(\Omega(T)/2)\},
\end{equation}
where $\Omega(T)= \hbar \omega / k_BT$ and $n(x)=1/(e^x-1)$ is the Bose-Einstein distribution function. $\omega_1(T)$ and $\Gamma_1(T)$ involve mainly the three-phonon decay process, in which an optical phonon decays into two acoustic modes with equal energies and opposite momenta. 
 \\
     
\section{Direct product of $M_{5}\otimes M_{5}$}~\label{Direct_product}    
In this appendix, we study the possibility of two-phonon Raman processes due to the $M_{5}$ modes. The combination of two $M_{5}$ phonons
with the same wavevector can couple with a number of modes from the
$\Gamma$ point, whereas the combination of two $M_{5}$ phonons from
different $M$ points can couple with phonons from the only other $M$
point in the Brillouin zone. The complete list of modes that two $M_{5}$
modes couple to can be predicted by taking the product $M_{5}$ irreducible representation (irrep)
by itself, $i.e.$ $M_{5}\otimes M_{5}$, and decomposing it into irreducible
representations of the space group. Since $M_{5}$ is a six-dimensional
irrep (it has a two-dimensional little group irrep on each of the three different $M$
points in the star), this product is 36 dimensional. Only 9 of these
36 degrees of freedom are at the $\Gamma$ point, and they are 
\begin{equation}
M_{5}\otimes M_{5}=\Gamma_{1}\oplus\Gamma_{3}\oplus\Gamma_{4}\oplus\Gamma_{5}+(\cdots),
\end{equation}
where the $(\cdots)$ includes irreps from the $M$ point. In other words,
the two-phonon processes from $M_{5}$ modes can couple with a $\Gamma_{1}$,
 $\Gamma_{3}$, $\Gamma_{4}$, or  $\Gamma_{5}$ phonon. (Using
the point group irreps labels, these are the $A_{1}$, $E$, $T_{1}$,
and $T_{2}$ modes, respectively.) 

The next question that is relevant is what combinations of the 6 degrees
of freedom of the $M_{5}$ modes couple with these modes. In the following,
for simplicity, we denote the six symmetry adapted components of $M_{5}(a,b;c,d;e,f)$
as $(M_{1a},M_{1b},M_{2a},M_{2b},M_{3a},M_{3b})$, respectively. If we denote
the only component of the fully symmetric irrep $\Gamma_{1}$ as $\Gamma_{1}$,
the coupling term that is second order in $M_{5}$ and first order
in $\Gamma_{1}$ ($A_{1}$) is 
\begin{equation}
\Gamma_{1}(M_{1a}^{2}+M_{1b}^{2}+M_{2a}^{2}+M_{2b}^{2}+M_{3a}^{2}+M_{3b}^{2}).
\end{equation}
 Similarly, the coupling between $\Gamma_{3}$ ($E$), whose components
we denote as $\Gamma_{3a}$ and $\Gamma_{3b}$, is 
\begin{equation}
\begin{split}
&\Gamma_{3a}(M_{1a}^{2}+M_{1b}^{2}-\frac{1}{2}[M_{2a}^{2}+M_{2b}^{2}+M_{3a}^{2}+M_{3b}^{2}])\\
&+\frac{\sqrt{3}}{2}\Gamma_{3b}(M_{2a}^{2}+M_{2b}^{2}-M_{3a}^{2}-M_{3b}^{2}).
\end{split}
\end{equation}
With a similar notation, the coupling to $\Gamma_{4}$ ($T_{1}$)
is 
\begin{equation}
\Gamma_{4a}M_{1a}M_{1b}+\Gamma_{4b}M_{2a}M_{2b}+\Gamma_{4c}M_{3a}M_{3b},
\end{equation}
 and the coupling to $\Gamma_{5}$ ($T_{2}$) is 
\begin{equation}
\Gamma_{5a}(M_{1a}^{2}-M_{1b}^{2})+\Gamma_{5b}(M_{2a}^{2}-M_{2b}^{2})+\Gamma_{5c}(M_{3a}^{2}-M_{3b}^{2}).
\end{equation}
                                                                                                                                                                                                                                                                                                                                                                                                                                                                                                                                  

%

\end{document}


\title{Supplemental Material for: \\
Phonon anomalies within the polar charge density wave phase
of the structurally chiral superconductor Mo$_3$Al$_2$C}

\author{Shangfei Wu}
\email{wusf@baqis.ac.cn}
\affiliation{Department of Physics and Astronomy, Rutgers University, Piscataway, New Jersey 08854, USA}
\affiliation{Beijing Academy of Quantum Information Sciences, Beijing 100193, China}
\author{Xianghan~Xu}
\affiliation{Department of Physics and Astronomy, Rutgers University, Piscataway, New Jersey 08854, USA}
\affiliation{Keck Center for Quantum Magnetism, Rutgers University, Piscataway, New Jersey 08854, USA}
\affiliation{School of Physics and Astronomy, University of Minnesota, Minneapolis, MN, USA}
\author{Fei-Ting~Huang}
\affiliation{Department of Physics and Astronomy, Rutgers University, Piscataway, New Jersey 08854, USA}
\affiliation{Keck Center for Quantum Magnetism, Rutgers University, Piscataway, New Jersey 08854, USA}
\author{Turan~Birol} 
\affiliation{Department of Chemical Engineering and Materials Science, University of Minnesota, MN 55455, USA}
\author{Sang-Wook~Cheong} 
\affiliation{Department of Physics and Astronomy, Rutgers University, Piscataway, New Jersey 08854, USA}
\affiliation{Keck Center for Quantum Magnetism, Rutgers University, Piscataway, New Jersey 08854, USA}
\author{Girsh~Blumberg} 
\email{girsh@physics.rutgers.edu}
\affiliation{Department of Physics and Astronomy, Rutgers University,
Piscataway, New Jersey 08854, USA}
\affiliation{National Institute of Chemical Physics and Biophysics,
12618 Tallinn, Estonia}
\date{\today}             
                                                                                            
\maketitle

 \section{Comparison between the as-grown and polished sample}\label{Compasion_of_sample}  

In this section, we show that polishing the sample does not affect the properties of Mo$_3$Al$_2$C  by comparing the linewidth between an as-grown (1 1 0) oriented and a polished (0 0 1) oriented Mo$_3$Al$_2$C.

\begin{figure*}[b] 
\begin{center}
\includegraphics[width=0.9\columnwidth]{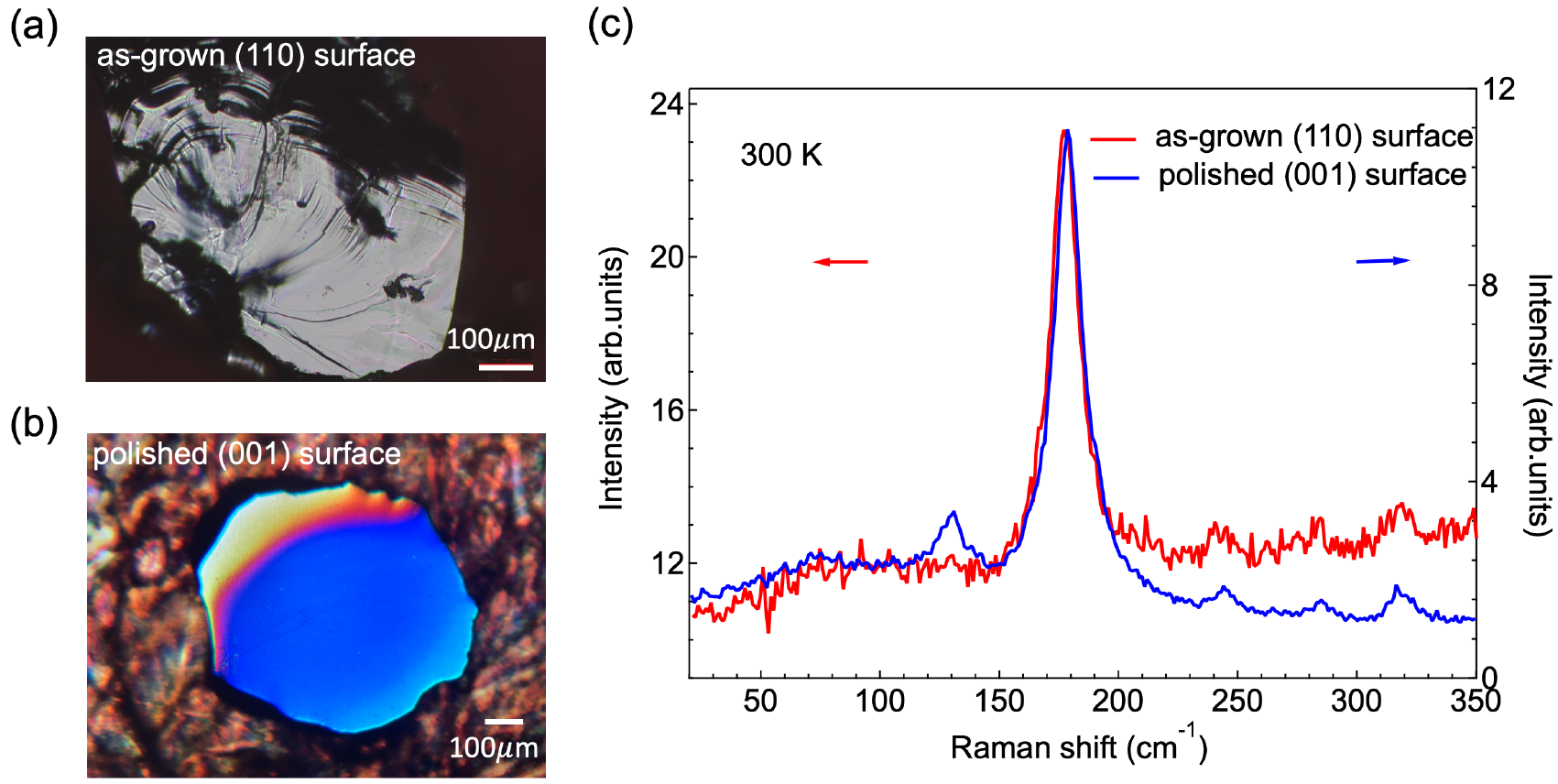}
\end{center}
\caption{\label{compare_asgrown_and_polished_sample}  
(a) Optical image of the top surface from an as-grown (1 1 0) orientated Mo$_3$Al$_2$C. 
(b) Polarized optical image of the top surface from a polished (0 0 1) orientated Mo$_3$Al$_2$C. 
(c) Raman spectra of the as-grown and polished Mo$_3$Al$_2$C in parallel scattering geometry.}
\end{figure*}

In Fig.~\ref{compare_asgrown_and_polished_sample}(a), we show the optical image of the top surface from an as-grown (1 1 0) orientated Mo$_3$Al$_2$C. In Fig.~\ref{compare_asgrown_and_polished_sample}(b), we present the polarized optical image of the top surface from a polished (0 0 1) orientated Mo$_3$Al$_2$C. The majority of the polished sample is in uniform color, suggesting that the polishing-induced strain or stress is negligible. We can always find a strain-free spot for our study, because the laser spot is about $50\times100$ $\mu$m$^2$ while the sample is around $600\times600$\,$\mu$m$^2$ in size. In Fig.~\ref{compare_asgrown_and_polished_sample}(c), we show two Raman spectra from these two samples in the parallel scattering geometry at 300\,K. The most intense phonon is the $A_1$ phonon located at around 180\,\cm-1. By comparing the Raman spectra at 300\,K, we find that polishing-induced linewidth broadening is negligible. Thus, a fine polishing of the sample does not affect the properties of the Mo$_3$Al$_2$C.

\section{Raman tensor analysis}\label{Raman_tensor_analysis}

The Raman tensor $R_{\mu}$ is a $2\times 2$ matrix for an irreducible representation ($\mu$) of a point group. It can be used to estimate the phonon intensity in a Raman scattering experiment.
With incident and scattering light polarization unit vector respectively defined as $\hat{e}_i$ and $\hat{e}_s$, the phononic Raman response is described as~\cite{Loudon_book}:
\begin{equation}
\label{eq_chi}
\chi''_{\hat{e}_i \hat{e}_s}\sim \sum_\mu|\hat{e}_i R_\mu \hat{e}_s |^2.
\end{equation}

The Raman tensors $R_{\mu}$ ($\mu$=$A_{1}$, $E$, $T_1$, $T_2$) for the irreducible representations ($\mu$) of point group $O$ have the following forms:
\begin{displaymath}   
R_{A_{1}}=\left(\begin{array}{ccc}
a & 0  &0\\
0 &a &0\\
0 & 0 &a
\end{array}\right),\\
R^1_E=\left(\begin{array}{ccc}
b & 0  &0\\
0 &b &0\\
0 & 0 &-2b
\end{array}\right), 
R^2_E=\left(\begin{array}{ccc}
-\sqrt{3}b & 0  &0\\
0 &\sqrt{3}b &0\\
0 & 0 &0
\end{array}\right)
\end{displaymath}  
\begin{displaymath}   
R^1_{T_1}=\left(\begin{array}{ccc}
0 & 0  &0\\
0 &0 &c\\
0 & -c &0
\end{array}\right),
R^2_{T_1}=\left(\begin{array}{ccc}
0 & 0  &c\\
0 &0 &0\\
-c & 0 &0
\end{array}\right),
R^3_{T_1}=\left(\begin{array}{ccc}
0 & c  &0\\
-c &0 &0\\
0 & 0 &0
\end{array}\right).
\end{displaymath}  
\begin{displaymath}   
R^1_{T_2}=\left(\begin{array}{ccc}
0 & 0  &0\\
0 &0 &d\\
0 & d &0
\end{array}\right),
R^2_{T_2}=\left(\begin{array}{ccc}
0 & 0  &d\\
0 &0 &0\\
d & 0 &0
\end{array}\right),
R^3_{T_2}=\left(\begin{array}{ccc}
0 & d  &0\\
d &0 &0\\
0 & 0 &0
\end{array}\right).
\end{displaymath}                                          
Note that $E$ irreducible representation is two-fold degenerate, while $T_1$ and $T_2$  irreducible representations are three-fold degenerate.

\textit{(0~0~1) surface}. We choose $\hat{e}_i$ and $\hat{e}_s$ to be $X$, $Y$, $R$, and $L$, where $X=(1~0~0)$, $Y=(0~1~0)$, $R=1/\sqrt{2} ~(1~i~0)$, and $L=1/\sqrt{2}~(1~-i~0)$, which they are defined on the (0~0~1) suface.

Based on Eq.~(\ref{eq_chi}), we obtain the Raman response in six scattering geometries: 
\begin{equation}
\label{scattering_geometryO1}
\begin{split}     
\chi''_{XX}&=a^2+4b^2,\\
\chi''_{XY}&=c^2+d^2,\\
\chi''_{X'X'}&=a^2+b^2+d^2,\\
\chi''_{X'Y'}&=3b^2+c^2,\\
\chi''_{RR}&=a^2+b^2+c^2,\\
\chi''_{RL}&=3b^2+d^2.\\
\end{split}
\end{equation}
Thus, the Raman selection rules for the $O$ point group indicate that the $XX$, $XY$, $X'X'$, $X'Y'$, $RR$, and $RL$ polarization geometries probe the $A_{1} + 4E$, $T_1 + T_2$, $A_{1} + E + T_2$, $3E+T_1$, $A_1+E+T_1$, and $3E+T_2$ symmetry excitations, respectively~[Table~I in the main text].

The sum rules  
\begin{equation}
\label{scattering_geometryO2}   
\chi''_{XX} + \chi''_{XY}=\chi''_{X'X'} + \chi''_{X'Y'} = \chi''_{RR} + \chi''_{RL} = a^2+4b^2+c^2+d^2,
\end{equation}
set a constraint for the Raman response in different scattering geometries, thus providing a unique way to check the data consistency.

Combining Eq.~(\ref{scattering_geometryO1}) and Eq.~(\ref{scattering_geometryO2}), we can calculate the square of the Raman tensor elements:
\begin{equation}
\label{decompossiitonD6hEQ2}
\begin{split}     
a^2&=(1/3)(\chi''_{XX}+\chi''_{X'X'}+\chi''_{RR}-\chi''_{X'Y'}-\chi''_{RL}),\\
b^2&=(1/6)(\chi''_{X'Y'}+\chi''_{RL}-\chi''_{XY}),\\
c^2&=(1/2)(\chi''_{XY}+\chi''_{RR}-\chi''_{X'X'}),\\
d^2&=(1/2)(\chi''_{XY}+\chi''_{RL}-\chi''_{X'Y'}).\\
\end{split}
\end{equation}

Therefore, the algebra in Eq.~(\ref{decompossiitonD6hEQ2}) can be used to decompose the measured Raman signal into four separate irreducible representations ($A_{1}$, $E$, $T_1$, $T_2$) of point group $O$~[Table~II in the main text].

\textit{(1~1~1) surface}. Following the same procedures above, we can obtain the selection rules on the (1~1~1) surface of Mo$_3$Al$_2$C [Table~III in the main text].

\section{Details of the Multi-Lorentzian fitting} \label{fitting} 
In this section, we show the details of the multi-Lorentzian fitting of the Raman peaks shown in Fig.5 of the main text.

The fitting function has the form:                                                                                                                                                
\begin{equation}
\label{decompossiitonD6hEQ}
\begin{split}     
f(\omega)&=\Sigma_{i} A_i \text{Lor}_i(\omega,\gamma_i,\omega_i)+\text{Bg}(\omega),\\
\end{split}
\end{equation}
where $\text{Lor}_i(\omega,\gamma_i,\omega_i)$ is the Lorentzian function for mode $i$, $A_i$ is the amplitude of the mode , $\omega_i$ is the peak position, $\gamma_i$ is the half width of half maximum, and $\text{Bg}(\omega)$ is the background function.
We use the anti-symmetrized Lorentzian function, which has the following form:
\begin{equation}
\label{decompossiitonD6hEQ}
\begin{split}     
\text{Lor}_i(\omega,\gamma_i,\omega_i)=\frac{4 \gamma_i  \omega  \text{$\omega_i $}}{\left(\gamma_i ^2+(\omega -\text{$\omega_i $})^2\right) \left(\gamma_i ^2+(\omega +\text{$\omega_i$})^2\right)}\\
\end{split}
\end{equation}
For the fitting Raman peaks in the $XX$ and $XY$ scattering geometries [Fig.5(a)-(d) and Fig.5(i)-(l) of the main text], a smooth empirical background function is used:
\begin{equation}
\label{decompossiitonD6hEQ}
\begin{split}     
\text{Bg}_1(\omega)=a*arctan(\omega /b),
\end{split}
\end{equation}
where $a$ and $b$ are adjustable parameters.

For the fitting of Raman peaks in the $X'Y'$ scattering geometries [Fig.5(e)-(h) of the main text], a linear background in the vicinity of each Raman mode is used.
\begin{equation}
\label{decompossiitonD6hEQ}
\begin{split}     
\text{Bg}_2(\omega)=c*\omega + d,
\end{split}
\end{equation}
where $c$ and $d$ are adjustable parameters.

\section{Temperature dependence of the superlattice Bragg peaks and polar domains imaging around $T^*$} \label{TEM}          
                                                                                                             
In this section, we show that superlattice Bragg peaks and polar domains  appear at around the same temperature $T^*$.     

Single crystal of Mo$_3$Al$_2$C was polished, followed by Ar-ion milling, and studied using a JEOL-2010F field-emission transmission electron microscopy (TEM) equipped cryogenic sample stage~\cite{Huang_2019_NC}.

\begin{figure*}[t] 
\begin{center}
\includegraphics[width=\columnwidth]{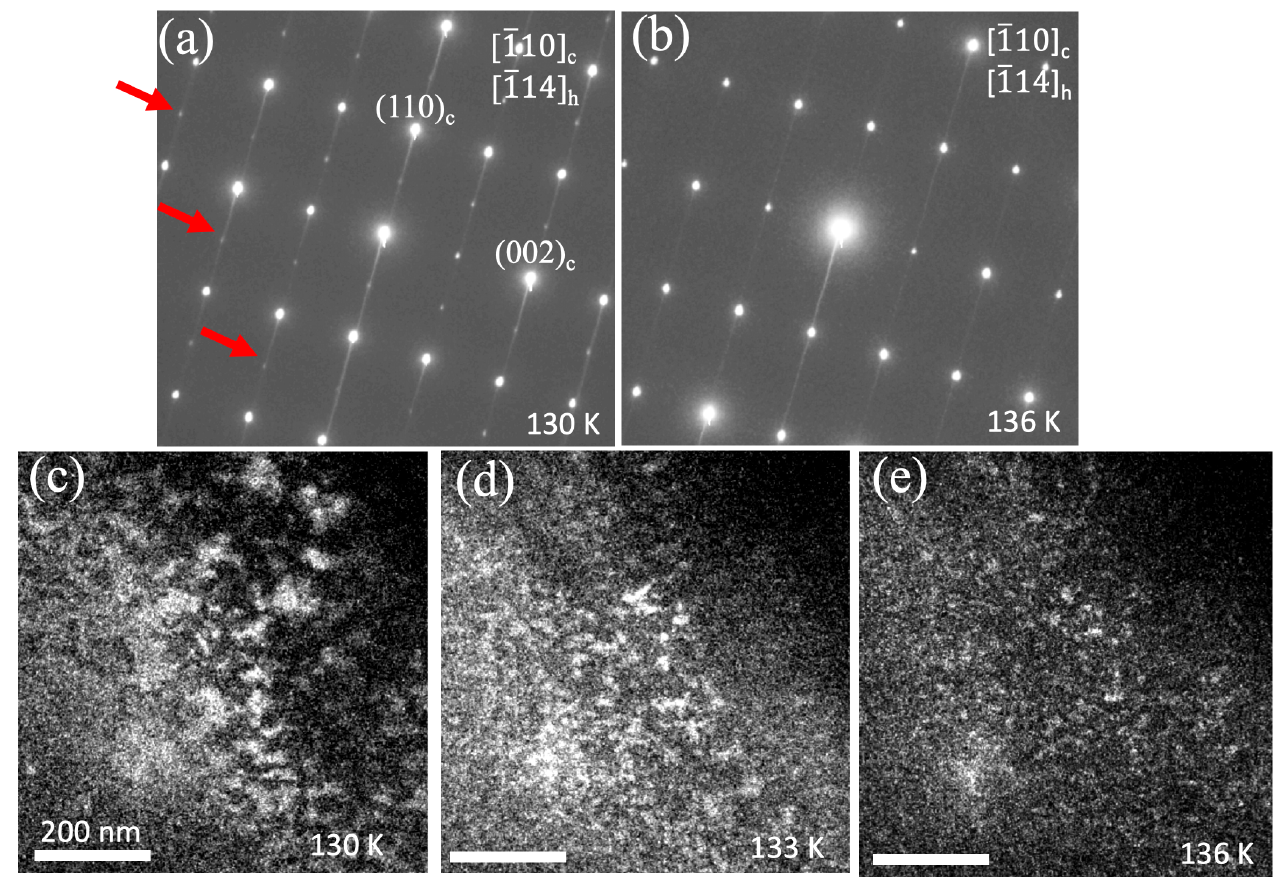}
\end{center}
\caption{\label{TEM}  
(a)-(b) Temperature-dependent selected area electron diffraction patterns (SAED) of Mo$_3$Al$_2$C single crystal along [$\bar{1}$~1~0]$_\text{c}$/[$\bar{1}$~$\bar{1}$~$\bar{4}$]$_\text{h}$ during a warming process. Red arrows indicate the superlattice peaks at 130 K, which disappear when the temperature exceeds 136\,K. The subscript $\text{c}$ and $\text{h}$ denote the cubic and hexagonal notations, respectively. 
(c)-(e) Dark-field TEM images were taken at 130\,K (c), 133\,K (d), and 136\,K (e) using $g_1^+= (1/2, -1/2, 4)_\text{c}$ spot along [$\bar{1}$~1~0]$_c$. The polar domains visible as black and white areas vanish above 136\,K, coinciding with the disappearance of superlattice peaks.  
The temperatures shown in this figure are the temperature of the cold-finger in the cryostat, which are not corrected for the thermal gradient.}
\end{figure*}                                                                                                     

In Figs.~\ref{TEM}(a)-(b), we show the temperature dependence of the selected area electron diffraction patterns of Mo$_3$Al$_2$C single crystal along [$\bar{1}$~1~0]$_\text{c}$/[$\bar{1}$~$\bar{1}$~$\bar{4}$]$_\text{h}$ during a warming-up process. Red arrows indicate the superlattice peaks at 130\,K, which disappear when the temperature exceeds 136\,K.
This indicates that the CDW transition appears at around cryostat temperature 130$\sim$136\,K.
We note that the CDW transition temperature $T^*$ is determined to be 155\,K from the transport measurement~\cite{Wu_2024_NC}. The comparison between the TEM and transport results suggests that there is a thermal gradient of about 20\,K in the TEM measurement of Mo$_3$Al$_2$C. The large thermal gradient might be because Mo$_3$Al$_2$C is a bad metal from the transport measurements (Fig.~1 in the main text). 

In Figs.~\ref{TEM}(c)-(e), we present the Dark-field TEM images taken at 130\,K (c), 133\,K (d), and 136\,K (e) using $g_1^+= (1/2, -1/2, 4)_\text{c}$ spot along [$\bar{1}$~1~0]$_c$ in a warming-up process. The polar domains visible as black and white areas vanish above 136\,K, coinciding with the disappearance of superlattice Bragg peaks.  
                                                                                                                                                                                
We thus conclude that superlattice Bragg peaks and polar domains appear at the same temperature $T^*$, namely, the CDW and polar transitions emerge at the same temperature.


%